\documentclass{sigchi}

\toappear{\small To appear in proceeding of \emph{WebSci'13}, May 1 -- 5, 2013, Paris, France.}
\usepackage{balance}  
\usepackage{graphics} 
\usepackage{times}    
\usepackage{url}      

\makeatletter
\def\url@leostyle{%
  \@ifundefined{selectfont}{\def\UrlFont{\sf}}{\def\UrlFont{\small\bf\ttfamily}}}
\makeatother
\def\pprw{8.5in}
\def\pprh{11in}

\usepackage[pdftex]{hyperref}
\hypersetup{
pdftitle={Petition growth and success rates on the UK No.~10 Downing Street Website},
pdfauthor={LaTeX},
pdfkeywords={WebScience, proceedings, archival format},
bookmarksnumbered,
pdfstartview={FitH},
colorlinks,
citecolor=black,
filecolor=black,
linkcolor=black,
urlcolor=black,
breaklinks=true,
}

\begin{document}

\title{Petition Growth and Success Rates\\ on the UK No.~10 Downing Street Website}

\numberofauthors{3}
\author{
  \alignauthor Scott A. Hale\\
    \affaddr{Oxford Internet Institute}\\
    \affaddr{University of Oxford}\\
    \email{scott.hale@oii.ox.ac.uk}
  \alignauthor Helen Margetts\\
    \affaddr{Oxford Internet Institute}\\
    \affaddr{University of Oxford}\\
    \email{helen.margetts@oii.ox.ac.uk}
  \alignauthor Taha Yasseri\\
    \affaddr{Oxford Internet Institute}\\
    \affaddr{University of Oxford}\\
    \email{taha.yasseri@oii.ox.ac.uk}
}


\maketitle

\begin{abstract}
Now that so much of collective action takes place online, web-generated data can further 
understanding of the mechanics of Internet-based mobilisation. This trace data offers 
social science researchers the potential for new forms of analysis, using real-time 
transactional data based on entire populations, rather than sample-based surveys of what 
people think they did or might do. This paper uses a `big data' approach to track the 
growth of over 8,000 petitions to the UK Government on the No.~10 Downing Street website 
for two years, analysing the rate of growth per day and testing the hypothesis that the 
distribution of daily change will be leptokurtic (rather than normal) as previous research 
on agenda setting would suggest. This hypothesis is confirmed, suggesting that 
Internet-based mobilisation is characterized by tipping points (or punctuated equilibria) 
and explaining some of the volatility in online collective action. We find also that most 
successful petitions grow quickly and that the number of signatures a petition receives 
on its first day is a significant factor in explaining the overall number of signatures
a petition receives during its lifetime. These findings have implications for the
strategies of those initiating petitions and the design of web sites with the aim of 
maximising citizen engagement with policy issues.
\end{abstract}




{\terms{Design; Human Factors; Measurement; Documentation}

\vfill
\section{Introduction}

Increasingly collective action is occurring online; most mobilisations include an online
element and some take place almost wholly online. All this web-based activity leaves a 
digital footprint, which allows researchers to generate  real-time transactional data of 
political behaviour. This kind of data represents a shift for social science research, 
with a move away from sample-based survey data about what people think they did or will 
do, to transactional data about what people actually did based on whole populations 
\cite{Lazer06022009,Conte2012}. Yet 
the mechanics of online collective action---and the extent to which the rise of the 
phenomenon challenges existing theories and understanding of collective action behaviour 
and leads to new models of contemporary political behaviour---remain under--analysed. We 
are still at the start of exploiting the great potential of `big data' to understand the 
mechanics of online collective action.

This paper analyses a `big data' set of all online petitions submitted to No.~10 Downing 
Street website from February 2009 to March~2011 (a screenshot of the webpage is shown in Figure \ref{fig:screenshot}). 
The petitions site was first launched in 
November~2006, and over the course of its lifetime it received more than 8 million 
signatures from over 5 million unique email addresses.\footnote{\url{http://www.mysociety.org/projects/no10-petitions-website/}} 
While live, the site allowed anyone to view petitions, and any user with a valid 
email address could create a new petition or sign an existing petition. The government 
promised an official response to all petitions receiving at least 500 signatures, 
providing an internal definition for the `success' of a petition. This paper analyses 
the growth of petitions and the distinctive characteristics of the mobilisation 
curves of successful and unsuccessful petitions. First, we provide some background 
on online collective action in general and e-petitions in particular. We use previous 
work on collective action and political attention to develop a hypothesis regarding the 
development of online mobilisations: that they will be characterised by long periods of
stasis and short periods of rapid change, leading to a leptokurtic distribution of daily 
change. Second, we outline the methods used to test this hypothesis and third, provide 
the results, followed by a discussion of the policy and design implications of the findings.

\begin{figure}[!t]
\centering
\includegraphics[width=0.9\columnwidth]{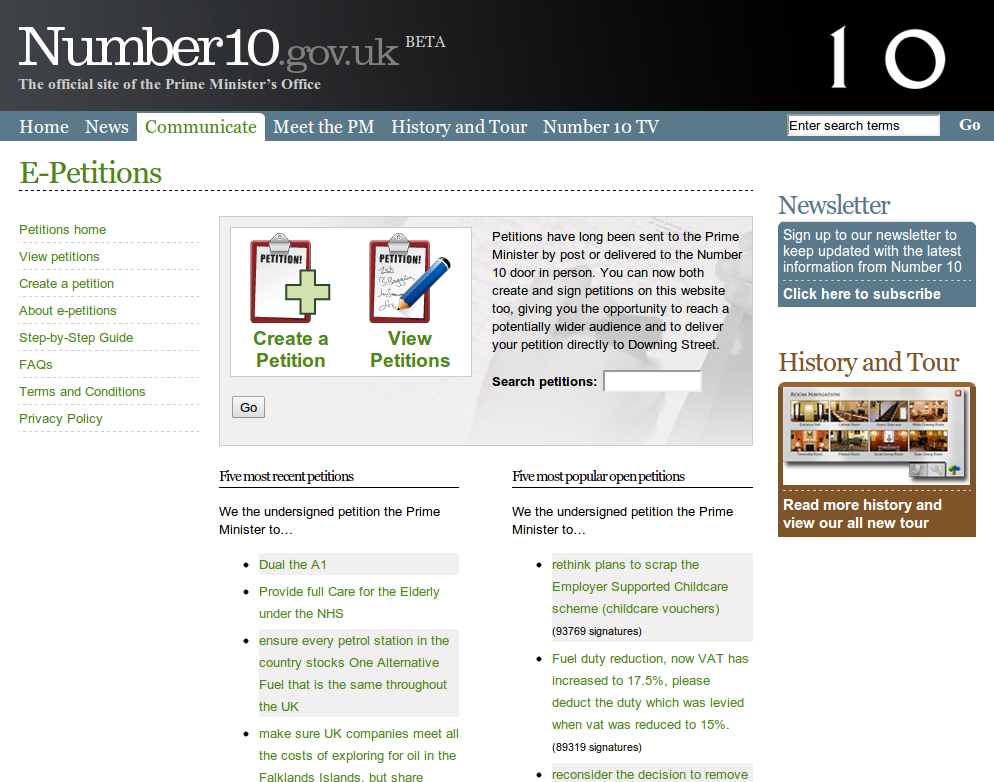}
\caption{No.~10 Downing Street petition website (no longer available)}
\label{fig:screenshot}
\end{figure}

\section{Background}
The 21st century has seen a rise in mobilisation, from the dramatic events in 
authoritarian states of the Arab Spring, to a series of protests, demonstrations 
and social backlash against austerity driven cutbacks and state retrenchment in 
liberal democracies facing the consequences of the financial crash of 2008. The 
web is implicated---to a larger or lesser extent---in virtually all these events. 
Researchers have turned their attention to the theoretical and conceptual 
implications of online collective action (e.g. \cite{Bimber2003,LupiaSin2003}) and 
some are using innovative methods, including experiments and data-mining, to explore 
the spread of mobilisations across online social networks (see, for 
example, \cite{AcklandGibson2006,EtlingEtAl2010,Hindman2008, GonzalezEtAl2011,BennettSegerberg2011,AralWalker2011}).

While online activity may be a minor element of some mobilisations, other mobilisations 
occur almost entirely online. The trend for `e-petitioning' represents one such activity, 
where online petitions are created, disseminated, circulated, and presented online, 
and although policy-makers may discuss responses in offline contexts, such responses 
are generated and sent online. The UK government's e-petition site was created on the 
No.~10 Downing Street website in November 2006 and ran until March 2011, when it was closed 
by the incoming Coalition government. Some of these petitions had high policy impact, 
notably one against the Labour administration's proposed road pricing policy, which 
policy-makers admitted off the record played a role in getting the policy scrapped. A 
new site was later launched in July 2011 on the direct.gov portal with a different format. 
Signing petitions has long been among the more popular political activities, leading the 
field for participatory acts outside voting and with other social benefits ascribed to 
it as well as having the potential to bring about policy change; e-petitioning 
reinforces `civic mindedness' \cite{WhyteEtAl2005}.

E-petitions are interesting examples of mobilisations with a strong online imprint, 
which will include the entire transaction history for both successful and unsuccessful 
mobilisations. The data generated from the signing of electronic petitions is an example 
of what is now commonly known as `big data', representing a transactional audit trail 
of what people actually did (as opposed to what people think they did) and an entire 
population (without the need to take a representative sample). Data like this represents 
a big shift for social science research into political behaviour, which has traditionally 
rested on survey data, or, for elections, voting data. Big data also presents 
challenges to social science research---it doesn't come with handy demographics 
attached and we do not know where people came from to any one interaction, nor where 
they are going, so it is often difficult to match up online activities across different 
platforms, or to identify the underlying factors influencing behaviour, such as age, 
income or gender. This data however, makes it possible to look at the different patterns 
of growth in the 8,000 mobilisation curves that we have and identify the distinctive 
characteristic of those mobilisations that succeed and those that fail with our digital 
hindsight. Such an analysis may tell us something about the nature of collective action 
itself in a digital world.

So what would we hypothesise about these mobilisations? A possible hypothesis may be 
derived from previous research on agenda setting in political systems. The most well 
known model of how policy attention proceeds in a liberal democracy is that of 
`punctuated equilibrium', developed by the US authors Baumgartner and Jones and their 
`Policy Agendas' programme of research (see \href{http://www.policyagendas.org/}{www.policyagendas.org}). 
The theory argues that policy attention to any issue will remain in long periods of stasis 
where little change occurs. Where issues do hit the policy agenda, it will be because some 
event has `punctuated' the equilibria, all eyes (including the media, public opinion, interest 
groups and politicians concerned) turn to the issue, money is spent, institutions are created 
and policy change occurs \cite{JohnMargetts2003,BaumgartnerJones1993,JonesBaumgartner2005}. 
The theory of punctuated equilibria is multi-faceted and has been illustrated by a range of 
empirical data across policy areas and within different dimensions of attention, such as 
public opinion, budgetary change and congressional attention \cite{JonesBaumgartner2005} 
and in various countries, including the UK \cite{JohnMargetts2003}. Baumgartner and Jones 
do not discuss Internet-related activity to any great degree; however, we might hypothesise 
that the pattern of mobilisations around a petition would proceed in a similar way, thereby
contributing to the same sort of issue attention cycle that has been observed many times over
in agenda setting research. Such a model would predict that the distribution of daily 
changes in attention would be `leptokurtic'.
Kurtosis is a measure of the ``peakedness'' of a distribution, and a distribution with 
positive excess kurtosis is said to be leptokurtic. Such distributions have an acute, 
slender peak around the mean and fatter tails. For the distribution of change in petitions 
data, this would correspond to a large amount of very little change (around the mean) with 
some punctuations of large change forming the tail of the distribution.

Such a finding could not show any causal effect, as only the activity of petitioners is being 
analysed here (and tipping points could suggest a media effect, although the authors are not 
aware of any petition having media coverage prior to having amassed a substantive amount of 
signatures). It could, however, point towards a role for online mobilisation in policy change 
analogous to that of the media in the agenda setting analysis, which is ascribed a lurching 
effect, due to the capacity of the media to parallel process only a small number of issues; 
at the point at which a punctuation occurs, media attention will `tip over' from specialist 
outlets into the mainstream media. In addition, in lower level mobilisations where the media
is not paying attention, such a finding could suggest a role in policy issues below the media radar.

\section{Methods}

The \href{http://web.archive.org/web/20100201044932/http://petitions.number10.gov.uk/}{UK Government's petition website}
was accessed daily from 2~February~2009 until March 2011, when the site closed, with an automated script. 
Each day, the number of overall signatures to date on each active petition was recorded. 
In addition, the name of the petition, the text of the petition, the launch date of the 
petition, and the category of the petition were recorded. Overall, 8,326 unique petitions 
were tracked, representing all publicly available petitions active at any point during the study.

Petitions on the Number 10 website closed, by default, 12 months after they first launched. 
To identify patterns in how petitions grow, the percentage change in new signatures was 
calculated each day. Most petitions had a long period of inactivity prior to their deadline 
date. To consider just how petitions grow, data was truncated after the last signature on a 
petition, removing any final period of zero signature-per-day growth prior to the petition's deadline.

Leptokurtic distributions have a more acute peak close to the mean and larger tails. 
There is no statistical test to specifically classify a distribution as leptokurtic. 
However, several tests in combination help demonstrate a distribution is leptokurtic 
(see \cite{JohnMargetts2003}). The most rigorous test is the Shapiro-Wilk test \cite{ShapiroWilk1965}, 
which checks whether the points could possibly be drawn randomly from a normal 
distribution. Leptokurtic distributions should reject the Shapiro-Wilk null hypothesis 
of normality. The Kolmogorov-Smirnov test tests that a set of frequencies is normal 
distributed by focusing on the skewedness and kurtosis of a 
distribution \cite[pp. 392--4]{ChakravartiEtAl1967}, and this null hypothesis should be 
rejected if a distribution is leptokurtic and hence non-normal. Visualizing the histogram 
and the quantile--quantile plot give further evidence of a leptokurtic distribution.

\section{Results}\label{sec:results}

The webcrawler collected all available petitions during the collection period, resulting 
in a set of 8,326 unique petitions. In Figure \ref{fig:zipf}, a Zipfian diagram shows the number of signatures at the end of data collection
 period vs rank of the petition according to the number of signatures. This demonstrates that most of the petitions receive only a few signatures, and only a few petitions (6 per cent) make it beyond the 500-signature threshold. The slop of the Zipf's law changes around the point of 500 signatures, indicating that the signatories are less eager to increase the number of signatures after meeting the number required for an official government response.

Nearly all of the 534 petitions that succeeded in 
obtaining 500 or more signatures did so quickly (Figure \ref{fig:growth}). Successful 
petitions took a mean time of 8.4 days to reach 500 signatures, but a median time of only 
two days. In fact, 230 of the 534 successful petitions succeeded in obtaining 500 
signatures on the day they were launched (day 1). Only a few petitions take a much longer 
time to reach the 500 signature mark: 31 petitions (6 per cent) succeed after taking more 
than 30 days, and only five successful petitions in our dataset required more than four 
months to reach the 500 signature mark.

Histograms of the logarithm of the number of signatures for petitions at different ages are shown in Figure \ref{fig:hist}. A fat-tailed
distribution emerges at the very start of the petitions, and as times passes only the lower bound of the distribution shifts slightly to larger numbers.
The long-term dynamic of petition growth is seen to have many small changes and a few very large changes.


\begin{figure}[!t]
\centering
\includegraphics[width=0.9\columnwidth]{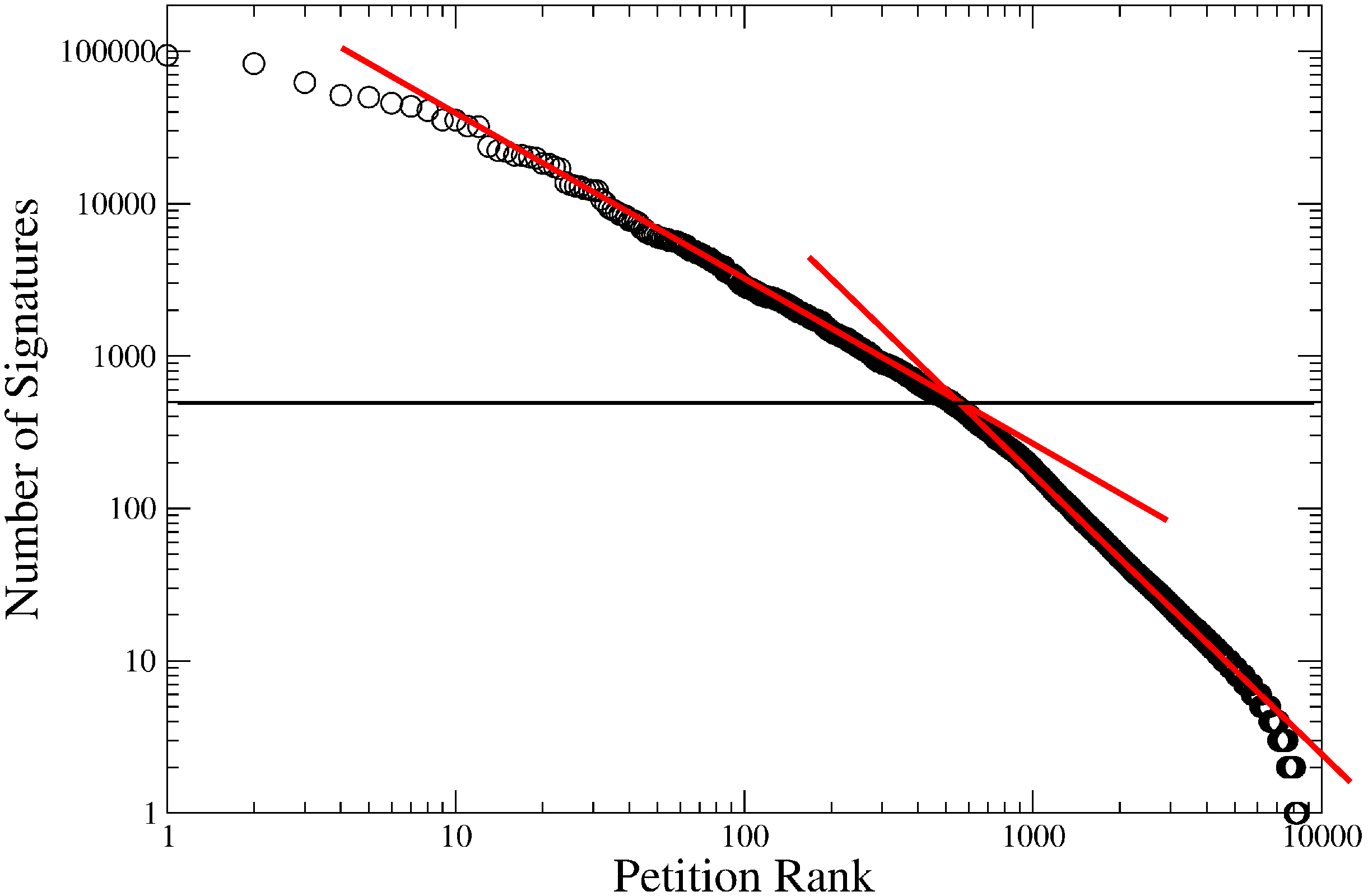}
\caption{Zipfian diagram of the petitions according to the final number of signatures received by each at the end of the data collection period. The change in the slope of the Zipf's law happens at 500 signatures suggesting a threshold effect (lines are to guide the eyes).}
\label{fig:zipf}
\end{figure}

\begin{figure}[!t]
\centering
\includegraphics[width=0.9\columnwidth]{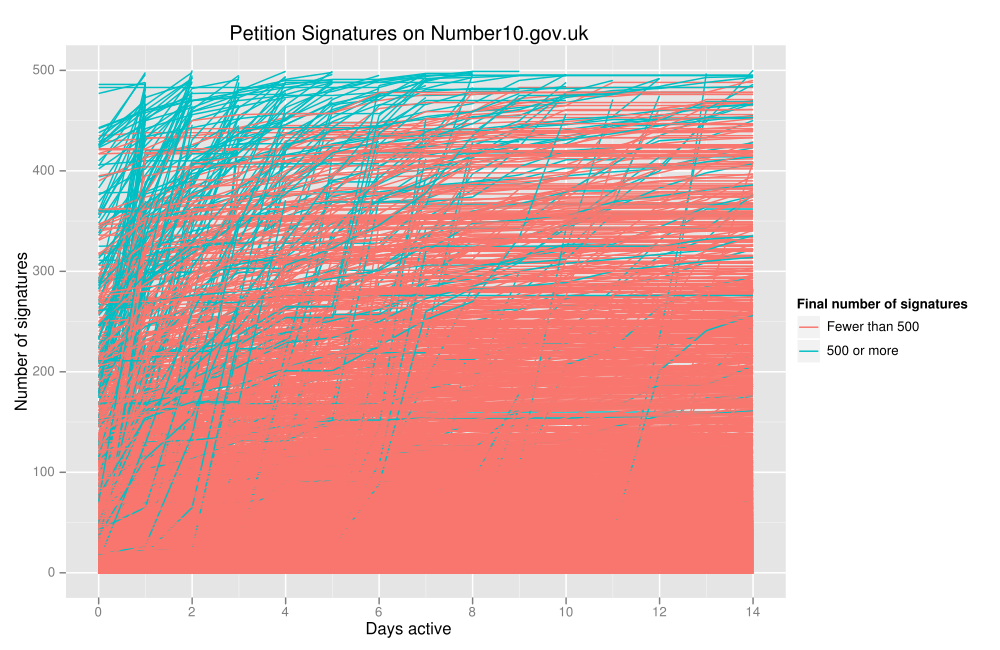}
\caption{Petition Growth, number of signatures received by each petition as a function of the number of days since its launch. The colour-coding
is according to the final number of signatures.}
\label{fig:growth}
\end{figure}

\begin{figure}[!t]
\centering
\includegraphics[width=0.9\columnwidth]{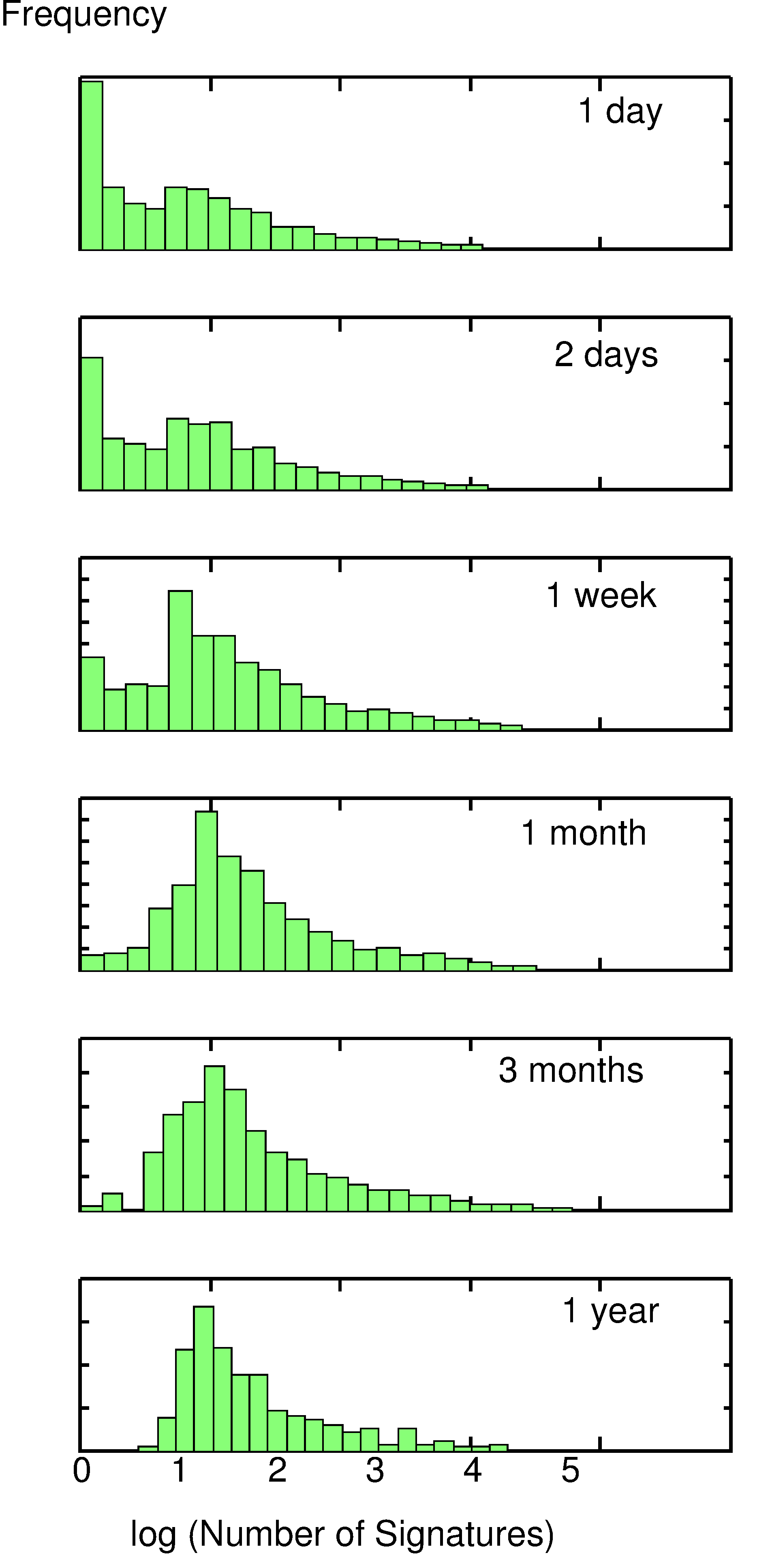}
\caption{Histogram of the logarithm of the number of signatures for samples of
petitions at different ages (indicated in each panel). The right tail of the distribution
emerges rapidly and as time increases, only the lower bound moves to right.}
\label{fig:hist}
\end{figure}

Figure \ref{fig:log_change} shows the percentage change in new signatures 
adjusted so that the mean growth of each petition lies at zero. While most daily 
change is small, petitions' growth is punctuated by a few large changes. The 
distribution of growth is leptokurtic and strongly rejects the Shapiro-Wilk null 
hypothesis of normality with a $W$ statistic of 0.17 translating to a p-value less 
than 0.000001. The distribution has a kurtosis score of 1,445 and a skewedness 
of 30.53, and rejects the Kolmogorov-Smirnov test for a normal distribution ($p<0.0001$). 
When we applied the same tests to the population of petitions that were successful in 
achieving 500 signatures (that is, excluding the unsuccessful ones), we found a similar 
leptokurtic distribution (Shapiro-Wilk $W$ statistic of 0.10, $p<0. 000001$).

\begin{figure}[!t]
\centering
\includegraphics[width=0.9\columnwidth]{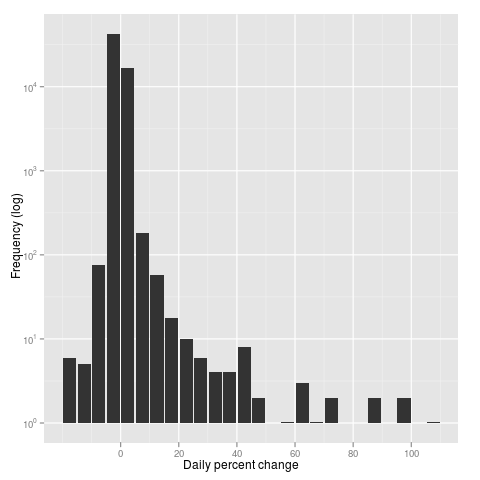}
\caption{Log of daily percentage change in number of signatories (centred around each petition's mean)}
\label{fig:log_change}
\end{figure}

The number of signatures a petition received on its first day has a significant, 
positive association to the log of the number of signatures the petition receives
in total during its lifetime (Table \ref{tbl:regression}). In addition, the number of 
other petitions started on the same day has a positive and significant regression coefficient.
The day of the week the petition launched on along with a simple weekend vs weekday 
launch distinction (not shown) were not significant. Certain categories of petitions 
were, however, associated with a greater or smaller number of signatures. In particular, petitions in
the Health, well-being and care category along with those in Environment
category received significantly more signatures, while petitions in the Government, politics and public administration category
received significantly fewer signatures. Both the Health and Environment categories are also significant in a
regression on the log of total signatures only including petitions receiving less than 500 signatures.
Overall, petitions tended to grow shortly after launch and then stop growing. This active period of
growth for petitions had a mean length of 57 days and a median length of 27 days.

\begin{table*}[tb]
  \centering
\begin{tabular}{l rrrr l}

  \hline
 & Estimate & Std. Error & t value & Pr($>$$|$t$|$) \\
  \hline
(Intercept) & 3.1253 & 0.0896 & 34.89 & 0.0000 \\
  cat: Economics and finance & -0.3360 & 0.0960 & -3.50 & 0.0005 & **\\
  cat: Education and skills & 0.0775 & 0.0984 & 0.79 & 0.4306 \\
  cat: Employment, jobs and careers & -0.2278 & 0.1067 & -2.14 & 0.0328 & * \\
  cat: Environment & 0.2603 & 0.0984 & 2.65 & 0.0082 & **\\
  cat: Government, politics and public administration & -0.2444 & 0.0895 & -2.73 & 0.0063 & **\\
  cat: Health, well-being and care & 0.2419 & 0.0860 & 2.81 & 0.0049 & **\\
  cat: Housing & -0.1822 & 0.1163 & -1.57 & 0.1173 \\
  cat: Information and communication & -0.0746 & 0.1269 & -0.59 & 0.5564 \\
  cat: International affairs and defence & 0.0739 & 0.1024 & 0.72 & 0.4705 \\
  cat: Leisure and culture & 0.2480 & 0.1084 & 2.29 & 0.0222 & *\\
  cat: Life in the community & 0.0002 & 0.1047 & 0.00 & 0.9983 \\
  cat: People and organisations & 0.1670 & 0.1208 & 1.38 & 0.1668 \\
  cat: Public order, justice and rights & 0.1279 & 0.0879 & 1.46 & 0.1457 \\
  cat: Science, technology and innovation & 0.3043 & 0.1654 & 1.84 & 0.0659 \\
  cat: Transport and infrastructure & -0.0669 & 0.0897 & -0.75 & 0.4561 \\
  Start on Monday & -0.0033 & 0.0586 & -0.06 & 0.9557 \\
  Start on Tuesday & -0.0461 & 0.0586 & -0.79 & 0.4315 \\
  Start on Wednesday & -0.0161 & 0.0591 & -0.27 & 0.7850 \\
  Start on Thursday & 0.0775 & 0.0595 & 1.30 & 0.1929 \\
  Start on Friday & -0.0651 & 0.0602 & -1.08 & 0.2793 \\
  Start on Saturday & -0.0153 & 0.0610 & -0.25 & 0.8022 \\
  Signatures on Day 1 & 0.0004 & 0.0000 & 28.31 & 0.0000 & ** \\
  Number of petitions started on same day & 0.0083 & 0.0020 & 4.11 & 0.0000 & **\\
   \hline
\end{tabular}
  \caption{Factors affecting growth. Ordinary least squares regression on the log of the total number of signatures\newline
  Categories variables are relative to the first category in the list, Business and industry.\newline
  ** $p<0.01$ ~~~ * $p<0.05$
  }
  \label{tbl:regression}
\end{table*}

\section{Discussion}
The 500 signatures mark seems at first consideration a very low threshold that 
should easily be passed. However, by far the majority of petitions (94 per cent in 
this time period) fail to attain even this modest number of signatures. This illustrates 
the low costs of initiating a collective action (with less regard to its viability) 
and the ability of digital trace data to easily track an entire population of 
mobilisations (before the success of each mobilisation is know). Petitions in the 
dataset are most active when they are first launched, and most petitions 
(presumably in the lack of outside stimulus) become digital dust after a couple 
of months despite typical deadlines of one year on the site.


Confirmation of our hypothesis regarding the leptokurtic distribution of changes 
to the support for a petition suggests that in online environments, collective 
action could play a role in a punctuated equilibrium model of policy change. That 
is, the general pattern for policy attention is for issues to remain dormant or in 
stasis, with a generally low level of attention. Some issues (by far the minority) 
that attract attention quickly gain a `critical mass' of activists and start to vie 
for policy attention, joining the range of other institutional influences in helping 
to `punctuate' the equilibrium. Such an argument would not include the claim that 
the mechanism by which collective action acts to bring about instability would be 
the same as the role played by the media, which plays a distinctive `lurching' role 
in Jones and Baumgartner's analysis, based in part on the tendency of the media to 
process a small number of issues in parallel. In the context studied here, the 
mechanism would depend more on the ways in which a mobilisation is disseminated 
via online social networks, something that previous research mentioned in the 
introduction has begun to investigate. If such activity tends to take place on 
the day the petition is initiated, then these findings could indicate the 
importance of achieving some kind of viability signal with petitioner's closest 
contacts right at the start, because initial rapid growth will have a greater 
effects on subsequent participation decisions by `weaker tie' contacts than a 
gradual growth over a prolonged period.

In other empirical studies using experimental methodologies, we have started to uncover 
the mechanism behind such punctuations. That is, in addition to the close contacts 
informed about the petition directly by the petitioner, other individuals (including 
people unknown to the petitioner or more distant contacts) deciding whether to 
participate will be influenced by the information that other people have already 
participated. This influence will depend on the number of other 
participants \cite{MargettsEtAl2011}, the personality of the individual deciding 
whether to participate \cite{MargettsEtAl-IP}, and the closeness of the individual 
to the petitioner. The occurrence of a punctuation will depend on the existence 
of `starters' whose thresholds for participation are low or whose closeness to the 
petitioner has in this instance reduced their threshold for participation. These 
starters will act as a signal for people with higher thresholds and weaker ties 
to the petitioner to `follow' in signing the petition, thereby acting as a further 
signal for people with even higher thresholds to join. At some point, if the petition 
is successful, then the number of followers will reach `critical mass' and attention 
to the mobilisation will become widespread, breaking out of the petitioner's social
network and gain more general social media exposure.

So, there are two possible explanations for the importance of the first day in achieving 
the critical mass: first, the way in which the petition is disseminated via online networks 
and second, the dynamic of starters and followers and social information about other 
participants in the burst of activity. If it is that visitors to the petition are heavily 
influenced by the numbers of other participants that they observe, then the finding could 
join those of other work in economics illustrating the importance of `first donations' 
in charitable giving \cite{BogEtAl2012}, showing that early donors set the precedent for 
later donors, or the wider literature on conditional co-operation, showing that social 
information about the contribution of others influences an individual participant's decision 
to contribute \cite{ShangCroson2009,FreyMeier2004}. This study complements this previous 
work by focusing on numbers of participants, rather than contribution amount (as everyone's 
contribution, at least that we are able to measure, is the same) and a more explicitly 
political context, rather than that of charitable giving.

The strong effect of petitions tending to succeed quickly or not at all may be influenced
by the design of the petition website during the period of study. For users starting at the 
homepage of the site, it was possible to view petitions overall or within a specific 
category and to sort petitions by the number of signatures or the date added. It was 
therefore easiest to look at petitions with the largest or smallest number of signatures 
and the oldest or newest petitions. It might be possible, therefore, that users of the 
site arriving at the homepage would only look at the newest petitions or the petitions 
with the most signatures contributing to the effects observed. This influence, however, 
would not apply to users following links shared via email and social media, which would 
point to a specific petition that the contact was supporting. These alternative effects 
might be tested using an experimental approach in future research.

\section{Conclusions}

We have found that in online mobilisations, growth tends not to occur, meaning that most 
mobilisations that are initiated, fail. But where it does, it proceeds in rapid bursts 
followed by periods of stasis. Such a finding suggests that online mobilisations of the 
kind covered here could play a role in the more general process of punctuated equilibria 
in policy-making. For example, Jones and Baumgartner \cite{JonesBaumgartner2005} found
a high correlation between public concern on an issue and Congressional attention. Our 
findings here could be even more interesting. In the theory of punctuated equilibrium the 
media plays a key role in terms of `lurching' from one issue to another and having a 
complex feedback relationship with public opinion. The sort of mobilisations we are 
looking at here, however, are bubbling up relatively independently of the media, gaining 
media attention only when they obtain significantly high levels of support---the 
petition on road-pricing that successfully played a role in obtaining policy change, 
for example, received a great deal of media attention once that it reached one million 
signatories.

Research that develops our understanding of the mechanics of this turbulence will be 
important for scholars and policy-makers alike as collective action continues to move 
into online settings. If online collective action is characterised by punctuations, 
then it looks as if such activity could inject a further dose of instability into 
political systems. If mobilisations follow a pattern of very low levels of attention 
punctuated by occasional `spurts' which grow rapidly into full scale mobilisations 
that merge with other elements of the political system to push policy change on to 
the agenda and the institutional landscape, then we can expect to see increasing 
turbulence in contemporary politics, adding to the `instability' that Baumgartner 
and Jones~\cite{BaumgartnerJones1993,JonesBaumgartner2005} and their co-investigators have modelled so extensively in
previous research.

Our future empirical work will analyse petition growth at a finer scale: we are currently
drawing data from a new UK e-petitions site on an hourly basis and from a US government 
petition website, in order to obtain a more fine-grained and cross-national analysis of 
the crucial early days of mobilisations. Future work will also explore coordination with media
coverage and mentions of the petitions on social networking sites (such as Facebook and 
Twitter), work to explicitly model petition growth curves, and better understand the role of petition categories.
Much of the work on collective action noted above is based on research into a
single platform, whereas any online activity tends to involve several. By looking 
carefully at the timing with which an issue gains attention in different parts of the 
political system, including the activist activities investigated here, we might get 
closer to establishing some sort of sequencing of attention. In addition, one way of 
getting around the lack of causal inference in research of this kind is to carry out 
experiments, as in Margetts et al.~\cite{MargettsEtAl2011,MargettsEtAl-IP}. The findings presented here seem to 
illustrate to those initiating petitions, the importance of disseminating a petition 
to their `strongest' contacts rapidly; future work using both `big data' and experimental 
approaches could further inform the optimal way that such dissemination should take place.

Future work will also use an experimental approach to analyse the effect of different 
information environments surrounding petition websites on petition growth. Designers of 
web sites that involve civic engagement, such as e-petition sites, must decide what 
social information about existing levels of participation to include, for example the 
numbers of people who have already signed and the timings of when they did so. Other 
design decisions include whether participants are anonymous or whether their names are 
made visible, and whether input from other social media platforms is incorporated into 
the petition site. Research of this kind can inform such design decisions in ways that 
maximise citizens' input to policy debates.

\section{Acknowledgements}
This work was supported by the Economic and Social Research Council [grant number RES-051-27-0331].

%
%
%
%
%
\balance


\end{document}